\documentclass[aps,prl,reprint,superscriptaddress,nobibnotes]{revtex4-1} 
\pdfoutput=1

\usepackage{graphicx}
\usepackage{amsmath,amssymb}
\usepackage{xspace}
\usepackage{siunitx}

\newcommand{\ket}[1]{\ensuremath{\vert #1 \rangle}\xspace}%
\begin{document}

\title{Observation of canted antiferromagnetism with ultracold fermions in an optical lattice}
\author{Peter T. Brown}
\affiliation{Department of Physics, Princeton University, Princeton, New Jersey 08544, USA}
\author{Debayan Mitra}
\affiliation{Department of Physics, Princeton University, Princeton, New Jersey 08544, USA}
\author{Elmer Guardado-Sanchez}
\affiliation{Department of Physics, Princeton University, Princeton, New Jersey 08544, USA}
\author{Peter Schau\ss}
\affiliation{Department of Physics, Princeton University, Princeton, New Jersey 08544, USA}
\author{Stanimir S. Kondov}
\affiliation{Department of Physics, Princeton University, Princeton, New Jersey 08544, USA}
\author{Ehsan Khatami}
\affiliation{Department of Physics and Astronomy, San Jos\'e State University, San Jos\'e, California 95192, USA}
\author{Thereza Paiva}
\affiliation{Instituto de Fisica, Universidade Federal do Rio de Janeiro, Caixa Postal 68.528, 21941-972 Rio de Janeiro RJ, Brazil}
\author{Nandini Trivedi}
\affiliation{Department of Physics, The Ohio State University, Columbus, OH 43210, USA}
\author{David  A. Huse}
\affiliation{Department of Physics, Princeton University, Princeton, New Jersey 08544, USA}
\author{Waseem S. Bakr}
\email{wbakr@princeton.edu}
\affiliation{Department of Physics, Princeton University, Princeton, New Jersey 08544, USA}


\begin{abstract}
Understanding the magnetic response of the normal state of the cuprates is considered a key piece in solving the puzzle of their high-temperature superconductivity \cite{Lee2006}. The essential physics of these materials is believed to be captured by the Fermi-Hubbard model \cite{Anderson1987}, a minimal model that has been realized with cold atoms in optical lattices \cite{Joerdens2008,Schneider2008}. Here we report on site-resolved measurements of the Fermi-Hubbard model in a spin-imbalanced atomic gas, allowing us to explore the response of the system to large effective magnetic fields. We observe short-range canted antiferromagnetism at half-filling with stronger spin correlations in the direction orthogonal to the magnetization, in contrast with the spin-balanced case where identical correlations are measured for any projection of the pseudospin. The rotational anisotropy of the spin correlators is found to increase with polarization and with distance between the spins. Away from half-filling, the polarization of the gas exhibits non-monotonic behavior with doping for strong interactions, resembling the behavior of the magnetic susceptibility in the cuprates \cite{Dagotto1994}. We compare our measurements to predictions from Determinantal Quantum Monte Carlo (DQMC) \cite{Blankenbecler1981} and Numerical Linked Cluster Expansion (NLCE) \cite{Rigol2006} algorithms and find good agreement. Calculations on the doped system are near the limits of these techniques, illustrating the value of cold atom quantum simulations for studying strongly-correlated materials.
\end{abstract}


\maketitle

Ultracold quantum gases have emerged as a powerful tool to study strongly correlated many-body physics. A two-component Fermi gas in an optical lattice can realize the repulsive Hubbard model, which describes  fermions in a periodic potential with onsite interaction $U$ and tunneling matrix element $t$ between neighboring sites \cite{Esslinger2010}. The recent introduction of quantum gas microscopes for fermionic atoms \cite{Cheuk2015,Parsons2015,Haller2015,Edge2015,Omran2015,Yamamoto2016,Greif2016} has led to rapid development in the experimental study of the 2D Hubbard model. The number-squeezed nature of the Mott insulating phase---previously inferred from bulk measurements \cite{Joerdens2008,Schneider2008}---has been explicitly revealed. Furthermore, site-resolved measurements probe antiferromagnetic correlations beyond the nearest neighbor \cite{Parsons2016,Boll2016,Cheuk2016}, which was not possible in previous studies \cite{Greif2013,Hart2015,Drewes2016b}.

\begin{figure*}[ht]
    \includegraphics[width=0.9\textwidth]{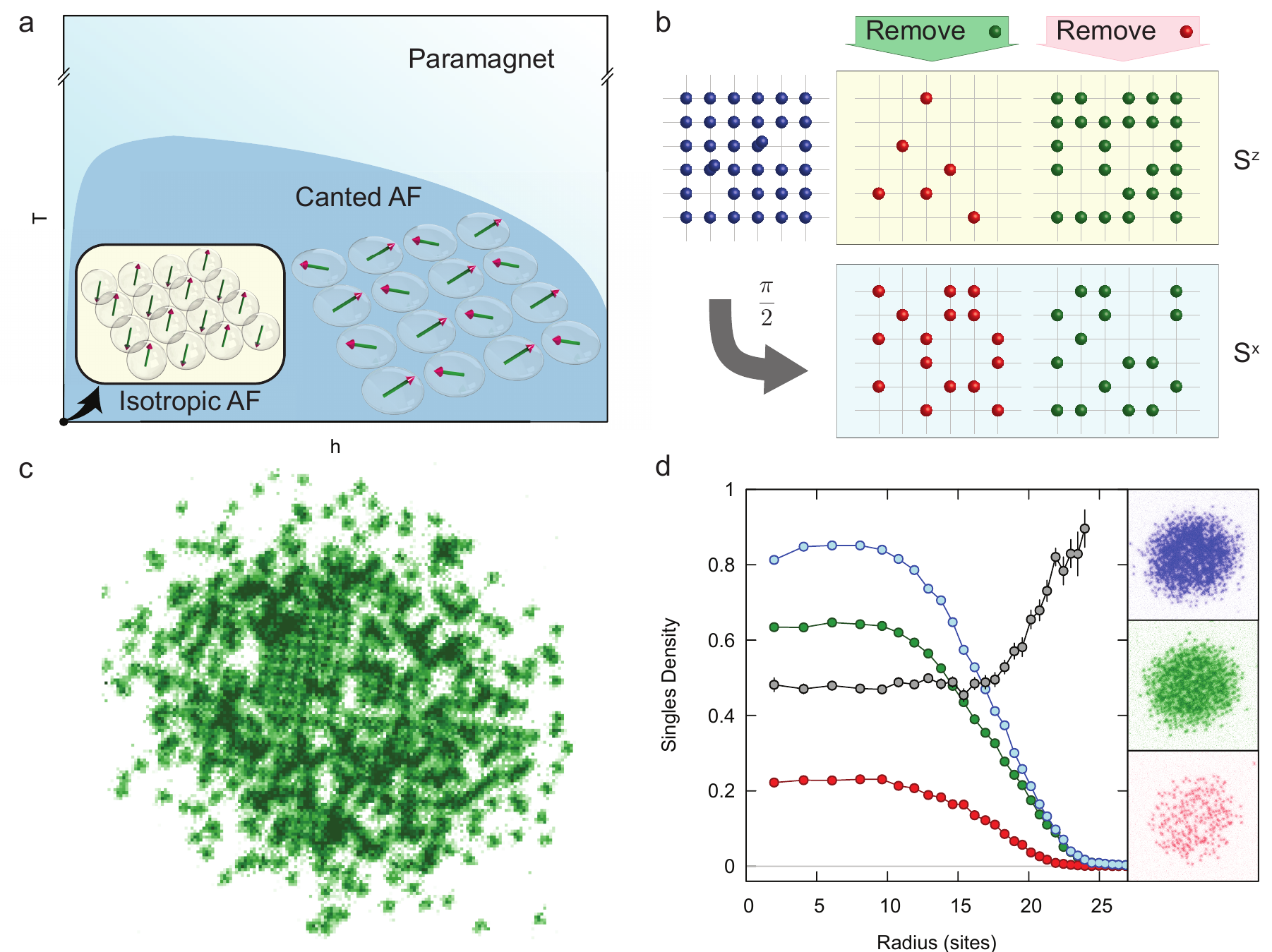}
    \caption{{\bf Site-resolved imaging of a spin-imbalanced Fermi gas in an optical lattice.} \textbf{a}, Schematic phase diagram of the Heisenberg model. $T$ is the temperature, and $h$ is the effective Zeeman field, controlled experimentally by the global polarization $P$. At $h=0$, the ground state is an antiferromagnet with SU(2) symmetry. For non-zero $h$, there is a finite temperature transition to a canted antiferromagnetic phase. Antiferromagnetic correlations also exist above the phase transition temperature, where they decay exponentially. The ellipsoids surrounding the spins illustrate the magnitude of correlations in a given direction. \textbf{b}, We prepare a spin mixture (blue) in an optical lattice, then selectively remove one spin state (red or green) and doublons. We extract spin correlations from the resulting density correlations for the $S^z$ spin projection and the $S^x$ projection after a spin rotation ($\pi/2$-pulse). \textbf{c}, Site-resolved fluorescence image after removal of one spin state. Field of view is $35\ \mu\text{m}$. \textbf{d}, Azimuthally averaged profiles and single fluorescence images showing $n^s_\uparrow$ (green), $n^s_\downarrow$ (red), $n^s$ (blue), and polarization $p^s$ (gray) over the trap with a spin-imbalanced Mott insulator at the center. Insets are exemplary single shot pictures. Field of view is $48\ \mu\text{m}$.}
    \label{fig:1}
  \end{figure*}

In this work, we investigate the Fermi-Hubbard model with imbalanced spin populations described by the Hamiltonian 
\begin{equation}
{\cal {H}}=-t\sum_{\langle ij\rangle,\sigma} \left(c^{\dag}_{i,\sigma}c_{j,\sigma}+c^{\dag}_{j,\sigma}c_{i,\sigma}\right) + U \sum_i n_{i,\uparrow}n_{i,\downarrow}.
\end{equation} Here $c^{\dag}_{i,\sigma}$ is the creation operator for a fermion with spin $\sigma$ on site $i$ and $n_{i,\sigma} = c^{\dag}_{i,\sigma}c_{i,\sigma}$. Theoretical studies of spin-imbalance in the Hubbard model have predicted an interesting magnetic structure in trapped gases arising from the interplay of spin-imbalance and antiferromagnetic and Stoner instabilities \cite{Koetsier2010,Wunsch2010,Snoek2011}. Experimentally, the polarization of our two-component atomic Fermi gas is a controllable quantity that is conserved due to the absence of spin-relaxation mechanisms. Thermodynamically, a non-zero polarization corresponds to the introduction of an effective Zeeman field $h=(\mu_{\uparrow}-\mu_{\downarrow})/2$, where $\mu_{\uparrow,\downarrow}$ are the chemical potentials of the two components. A starting point for understanding the  low temperature phase diagram at half-filling and strong interactions is the Heisenberg Hamiltonian \begin{equation}{\cal {H}}=J\sum_{\langle ij\rangle} {\bf S}_i\cdot {\bf S}_j - 2h\sum_i S^z_i,\end{equation} which is a good approximation for the Hubbard model in this regime. Here, ${\bf S}$ is the vector spin operator, $S^z = \pm \frac{1}{2}$ is its projection along the direction of the magnetization and $J=4t^2/U$ is the superexchange coupling. At zero field, the SU(2) symmetry of the Hamiltonian leads to isotropic antiferromagnetic correlations that decay exponentially in 2D with a correlation length that diverges at zero temperature. For a non-zero field, the Heisenberg model has a finite temperature Kosterlitz-Thouless transition to a canted antiferromagnet \cite{Koetsier2010} (Fig.~\ref{fig:1}a), which accommodates magnetization along the field while still benefiting from the superexchange interactions by building up long-range antiferromagnetic correlations of the spin components perpendicular to the magnetization. While the Heisenberg model provides useful insight in this regime, with finite $U/t$ or with doping it is necessary to consider the full Hubbard model.

  \begin{figure}[h]
   \includegraphics[width=1\columnwidth]{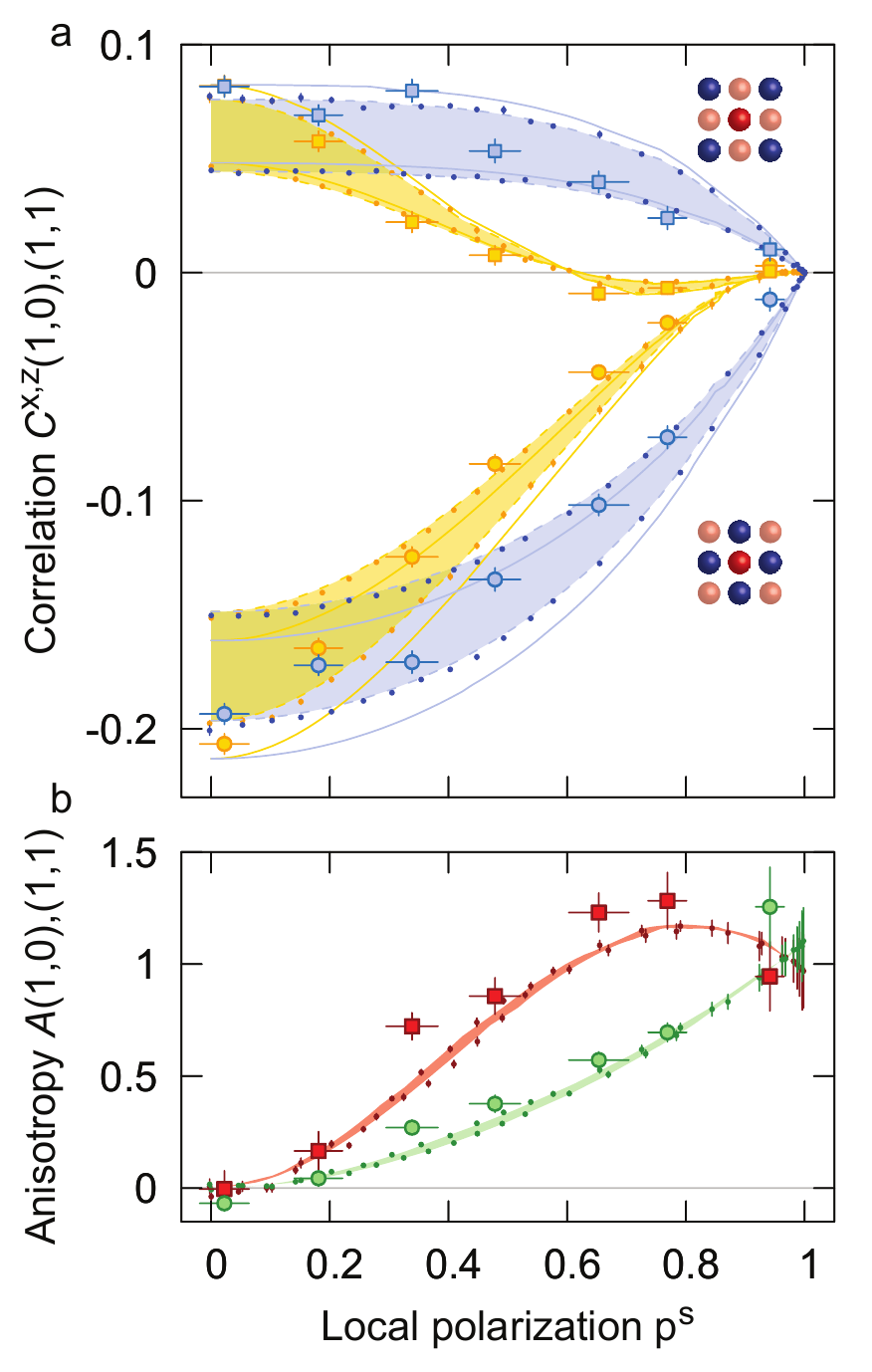}
    \caption{ {\bf Anisotropic spin correlations vs. polarization.}  \textbf{a}, Nearest neighbor (large circles) and diagonal neighbor (squares) spin correlations for the $S^z$ (yellow) and $S^x$ (blue) spin projections versus local polarization $p^s$ at half-filling. We show NLCE (dashed lines) and DQMC (small circles) results at $U/t=8$ corrected for our detection efficiency of 0.96 (see Methods) and uncorrected NLCE results (solid lines). We display a temperature band from $T/t = 0.38$ to $0.53$. \textbf{b}, Anisotropy $A$ of nearest-neighbor (large green circles) and diagonal neighbor (red squares) spin correlations with NLCE (solid lines) and DQMC (small circles) results. $A$ is insensitive to detection efficiency. Error bars are s.e.m. Experimental data averaged over $\sim50$ images and azimuthally.}
    \label{fig:2}
  \end{figure}

We realize the two-dimensional Fermi Hubbard model using a degenerate mixture of two hyperfine ground states $\left| \uparrow\right>$ and $\left|\downarrow\right>$ of $^6$Li in an optical lattice. The global spin imbalance $P=(N_\uparrow-N_\downarrow)/(N_\uparrow+N_\downarrow)$ can be varied continuously from 0 to $\approx0.9$ by evaporating the gas in the presence of a magnetic gradient leading to preferential loss of one of the spin states. We work at a scattering length of 448(9)~$a_0$, where $a_0$ is the Bohr radius, obtained by adjusting a bias magnetic field in the vicinity of a broad Feshbach resonance centered at 690~G. The imbalanced mixture is prepared in a single layer (for details see \cite{Mitra2016}) and subsequently loaded adiabatically into a 2D square lattice of depth $10.5(3)~E_R$, where $t = h\times 450(25)$~Hz. Here, $E_R=h\times14.66$~kHz is the recoil energy. For these parameters, we obtain $U/t=8.0(5)$, where strong antiferromagnetic correlations are expected at half-filling in the balanced gas. 
  
Fluorescence images obtained with quantum gas microscopy techniques enable us to identify singly occupied sites in the lattice, regardless of the spin state (see Methods). Doubly occupied sites undergo light assisted collisions and appear empty. We can also identify singly occupied sites where the atoms are projected onto a chosen eigenstate of $S^z$ by illuminating the cloud with a short pulse of resonant light that ejects atoms in the other eigenstate (Fig.~\ref{fig:1}b,c). By first converting atoms on doubly occupied sites to deeply bound molecules, we ensure that they are not affected by this light pulse and they are subsequently lost in light-assisted collisions during imaging. From these images we extract the azimuthally-averaged density of atoms on singly-occupied ($s$) sites in a particular spin ($\sigma$) state $n^s_{\sigma}(\mathbf{r})$ and the total density $n^s = n^s_\uparrow + n^s_\downarrow$, shown in Fig.~\ref{fig:1}d for an imbalanced gas. We observe a plateau in $n^s$ over an extended region near the center of the trap, indicating the formation of a spin-imbalanced Mott-insulating state. The deviation of $n^s$ from unity within the Mott insulator is primarily due to doublon-hole quantum fluctuations, which are non-negligible at our interaction strength. We characterize the local polarization in terms of the measured quantities $p^s=(n^s_\uparrow-n^s_\downarrow)/(n^s_\uparrow+n^s_\downarrow)$. This definition coincides with the true polarization in the absence of doubly occupied sites. At the accessible temperatures, the local polarization is constant throughout the Mott insulator.

We measure the spin correlators in the Mott insulator both for the spin components along the field, $C^z$, and orthogonal to it, $C^x$, by extending techniques previously introduced in experiments on balanced gases \cite{Parsons2016,Boll2016,Cheuk2016}. The spin correlator at a displacement $\mathbf{d}$ between two sites is given by $C^{\alpha}(\mathbf{d}) =  4\left(\langle S^{\alpha}_{\mathbf{i}} S^{\alpha}_{\mathbf{i}+\mathbf{d}} \rangle - \langle S^{\alpha}_{\mathbf{i}} \rangle\langle S^{\alpha}_{\mathbf{i}+\mathbf{d}}\rangle \right)$, where the brackets denote an ensemble and azimuthal average. $C^z(\mathbf{d})$ is obtained from the singles density correlators $\langle n^s_\mathbf{i} n^s_{\mathbf{i}+\mathbf{d}}\rangle$ and $\langle n^s_{\mathbf{i},\sigma}n^s_{\mathbf{i}+\mathbf{d},\sigma}\rangle$ taking into account the effect of doublons and holes (see Methods). To extract $C^x(\mathbf{d})$, we insert an additional radiofrequency pulse to rotate the spins by $\pi/2$ before initiating the measurement protocol.

The measured nearest and next-nearest neighbor correlators $C^{x,z}$  are shown versus $p^s$ in Fig.~\ref{fig:2}a. As the polarization is increased, we observe an overall decrease in correlations, with $C^z$ decreasing faster than $C^x$. We define the correlator anisotropy as $A=1-C^z/C^x$. For an almost unpolarized gas $p^s=0.02(4)$ we measure $A = -0.06(7)$ for the nearest neighbor and $A = 0.0(2)$ for the next-nearest neighbor. The consistency of these values with zero verifies the SU(2) symmetry of the Hubbard Hamiltonian at $h=0$. The anisotropies are shown versus $p^s$ in Fig.~\ref{fig:2}b. The system's preference to build correlations in the plane orthogonal to the field can already be understood at the level of a classical Heisenberg model, since spins oriented with the staggered magnetization in $xy$ the plane can lower their energy by uniformly canting in the direction of the field. In the quantum system, strong quantum fluctuations in 2D reduce the magnitude of the nearest-neighbor correlator in the balanced gas from 1 to 0.36 in the ground state \cite{Parsons2016}, and thermal fluctuations and imaging fidelity further reduce it to the experimentally measured value of $0.207(4)$. At non-zero polarization, we observe that the correlator anisotropy is stronger when the sites are further apart. For example, at $p^s=0.48(4)$, $A=0.38(9)$ for the nearest neighbor, while $A = 0.8(2)$ for the next-nearest neighbor. The increase of the correlation anisotropy with distance can be partly understood by considering what happens at lower temperatures as we approach the Kosterlitz-Thouless transition. There the $C^x$ correlations become long range while the $C^z$ do not, so at long distance $A$ approaches one.

\begin{figure*}[ht]
   \includegraphics[width=\textwidth]{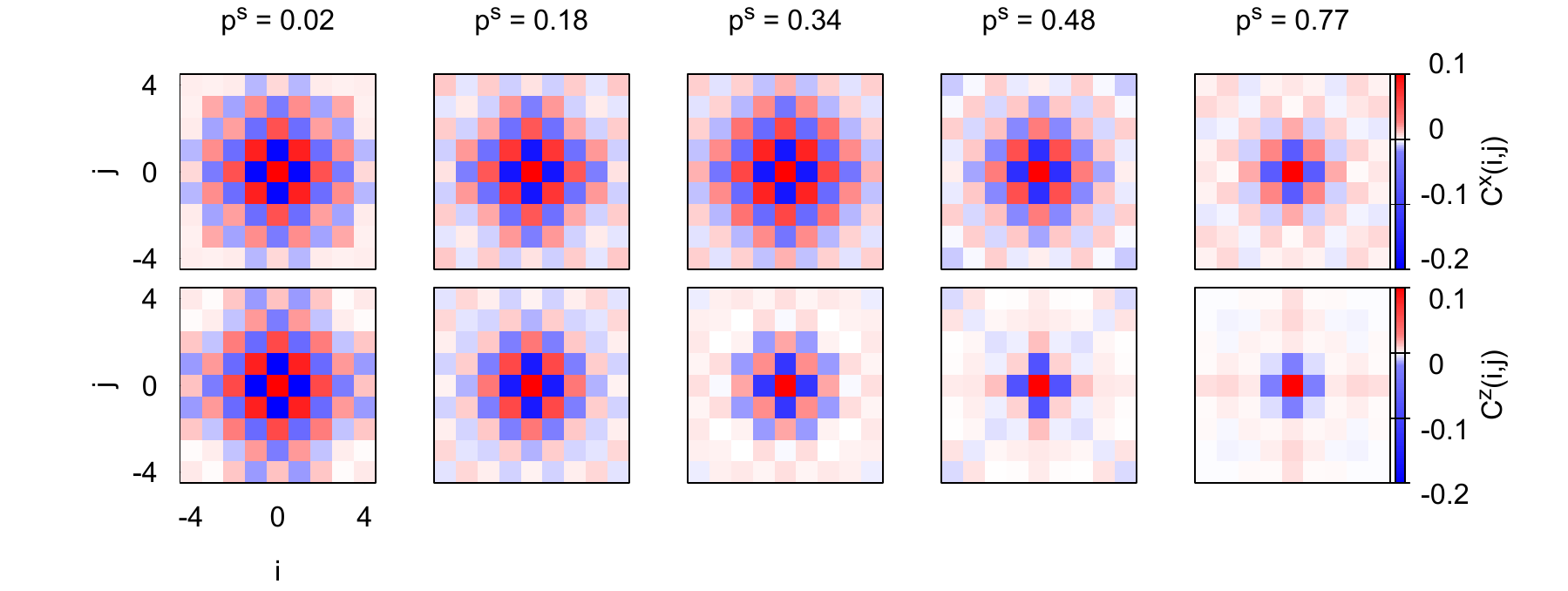}
    \caption{ {\bf Spin correlation matrices.}  Full spin correlation matrices for different site displacements, shown at half-filling for different polarizations $p^s$. Top row shows $S^x$ correlators, $C^x$, and bottom row shows $S^z$ correlators, $C^z$. Correlator values are averaged over symmetric points. See Methods for comparison with NLCE data.}
    \label{fig:3}
  \end{figure*}

Insight into the range of the antiferromagnetic order can be gained by examining 2D plots of the spin correlators as a function of the displacement vector between the sites, shown in Fig.~\ref{fig:3}. The checkerboard pattern is visible for displacements of up to four sites along the diagonal in the almost unpolarized gas, and the overall decrease of all correlations with polarization, as well as the relative suppression of $C^z$ relative to $C^x$ is evident. The $C^x$ correlations remain antiferromagnetic at all polarizations, but the $C^z$ correlations can be viewed as the density correlations of the gas of minority spins \cite{Batyev1984} whose modulation becomes longer wavelength as the density of this gas decreases. This leads to a change in the sign of $C^z(1,1)$ near $p^s = 0.6$ which can be seen in Figs. ~\ref{fig:2}a and ~\ref{fig:3}. The observation of this percent-level negative correlation is only possible because of the superb sensitivity of quantum gas microscopy.

We compare our measurements to results from DQMC and NLCE simulations of the Fermi-Hubbard model \cite{Paiva2010,Khatami2011} at half-filling in the presence of a chemical potential imbalance with the temperature left as a free parameter. For the balanced gas, the measured nearest neighbor correlators give a temperature of  $T/t = 0.40(5)$. The temperature shows a mild increase with polarization, rising to $T/t = 0.57(5)$ at $p^s=0.77(3)$. The calculated anisotropy is almost independent of temperature over this range, and shows excellent agreement with the experiment (Fig.~\ref{fig:2}b).

\begin{figure}[hb]
   \includegraphics[width=1\columnwidth]{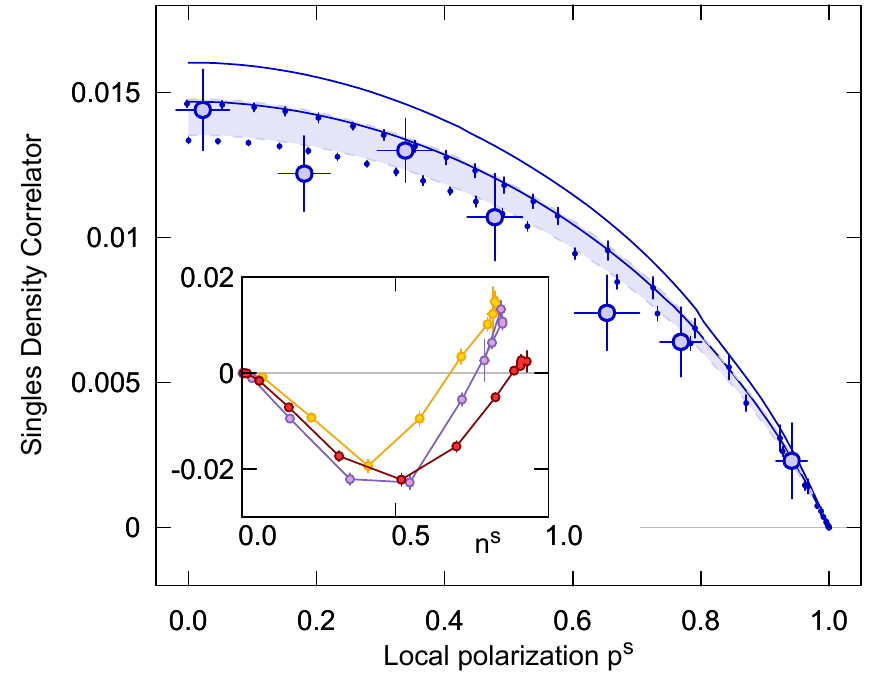}
    \caption{ {\bf Singles density correlations.}   Nearest-neighbor singles density correlations at half-filling versus singles polarization $p^s$. We show NLCE (dashed lines) and DQMC (small circles) results corrected for our detection efficiency, and uncorrected NLCE results (solid lines). We show a temperature band from $T/t = 0.38$ to $0.53$, which is determined from spin correlators. \textbf{Inset}, Singles density correlations versus density for $p^s = 0.02$ (yellow), $p^s = 0.48$ (purple), and $p^s = 0.94$ (red). Lines are guides to the eye. Error bars are s.e.m. Experimental data averaged over $\sim50$ images and azimuthally.}
    \label{fig:4}
  \end{figure}  

In addition to spin correlations, our data gives access to density correlations between singly occupied sites $\langle n^s_\mathbf{i} n^s_{\mathbf{i}+\mathbf{d}}\rangle_c = \langle n^s_\mathbf{i} n^s_{\mathbf{i}+\mathbf{d}} \rangle - \langle n^s_\mathbf{i}\rangle \langle n^s_{\mathbf{i}+\mathbf{d}}\rangle$ (Fig.~\ref{fig:4}). Previous analysis of this correlator in balanced gases revealed a dominant positive contribution at half-filling from doublon-hole virtual excitations and a smaller negative contribution from hole-hole correlations due to Pauli repulsion~\cite{Cheuk2016}. For increasing polarization, singles density correlations decrease as Pauli blocking suppresses double occupancy in the gas. The results of NLCE and DQMC calculations at $U/t=8$ show good agreement with the data using the same temperature range extracted from the spin correlations. In the inset of Fig.~\ref{fig:4}, we show the singles density correlation versus $\langle n^s\rangle$ for different polarizations. As was observed in the balanced case, the correlator changes sign as the filling is reduced, an effect that has been attributed to Pauli repulsion in the metallic regime \cite{Cheuk2016}. This repulsion becomes more pronounced as the polarization is increased, leading to negative correlations over a wider range of fillings.

   \begin{figure*}[ht]
    \includegraphics[width = \textwidth]{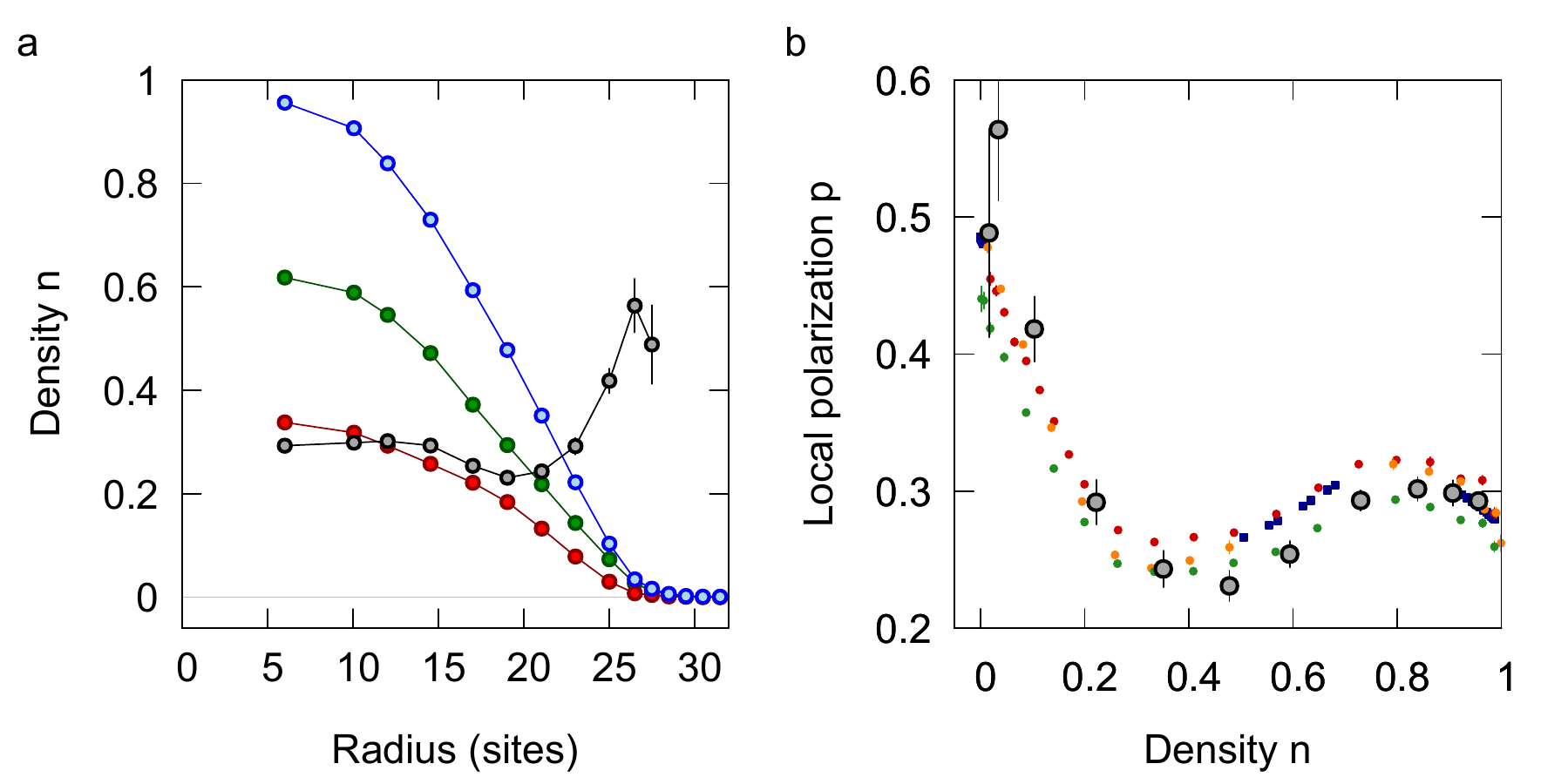}
    \caption{{\bf Polarization vs. doping at large interactions.} \textbf{a},  Azimuthally averaged profiles showing $n_{\uparrow}$ (green), $n_{\downarrow}$ (red), $n$ (blue), and $p$ (gray) for a spin-imbalanced Mott insulator at $U/t=14.7(8)$ and global polarization $P = 0.29(3)$. \textbf{b}, Local polarization as a function of density (gray circles). NLCE results (blue squares) at $U/t=14$ for $T/t = 0.42$ and $h/t = 0.22$. DQMC results (small circles) at $U/t=15$ for $T/t = 0.42$ and $h/t=0.22$ (red), $T/t = 0.36$ and $h/t = 0.20$ (orange) and $T/t = 0.42$ and $h/t = 0.20$ (green). Error bars are s.e.m. Experimental data averaged over $\sim55$ images and azimuthally.}
    \label{fig:5}
  \end{figure*}

The polarization profile of the imbalanced gas in the trap gives insight into the spin susceptibility of the Hubbard model in the doped regime. For strong interactions we observe that the in-trap polarization profile can exhibit non-monotonic behavior as shown in Fig.~\ref{fig:5}a for $U/t=14.7(8)$ obtained by increasing the scattering length to 793(12)~$a_0$. For these experiments, we extract the true polarization $p=(n_{\uparrow}-n_{\downarrow})/(n_{\uparrow}+n_{\downarrow})$, rather than $p^s$ (see Methods). The local polarization shows a shallow rise near the edge of the Mott insulator, then drops in the metallic region, before rising rapidly at the edge of the cloud. These effects can be understood qualitatively in terms of the magnetic susceptibilities of the gas at different fillings. At half-filling ($n=1$), the susceptibility is expected to be that of an antiferromagnet $\chi_{AF}  \propto 1/J = U/4t^2$, while for small doping at our temperatures there is a nondegenerate gas of holes in the lower Hubbard band and as a result a weak maximum in the susceptibility. At intermediate hole doping, the susceptibility crosses over to that of a metal, $\chi_{m} \propto 1/t$, which is smaller than $\chi_{AF}$ for large $U/t$. Similar behavior has been observed in the cuprates in the normal phase \cite{Torrance1989,Johnston1989,Takagi1989} and studied theoretically \cite{Glenister1993,Moreo1993}. At even lower filling, $T/T_F>1$ and there is no filled Fermi sea to hinder spins from aligning with the effective field, leading to an enhanced magnetic susceptibility. We show the polarization versus density in Fig.~\ref{fig:5}b. and compare to NLCE and DQMC calculations in the local density approximation.  The data is at a constant effective field since the trapped gas is in chemical equilibrium. The strength of the field, $h/t=0.21(1)$, is extracted from the experimental polarization at half-filling, which exhibits only a weak dependence on temperatures for $T/t<0.5$. These calculations reproduce the non-monotonic behavior of the polarization.

We have presented the first study of a 2D spin-imbalanced Fermi-Hubbard system in a regime near the edge of what state-of-the-art numerical techniques can simulate. Our quantum simulations qualitatively reproduce the non-monotonic behavior of the magnetic susceptibility with doping in the cuprates. On the other hand, the high effective  fields reached in our experiments allow us to explore canted antiferromagnetism, physics that is inaccessible in the cuprates at laboratory magnetic fields. Interesting future directions for both experimental and theoretical work include investigation of spin-imbalance in the attractive 2D Hubbard model where Fulde-Ferrell-Larkin-Ovchinnikov superfluid correlations should be detectable at the entropies achieved in repulsive experiments \cite{Gukelberger2016} and study of the Kosterlitz-Thouless transition in the imbalanced repulsive gas \cite{Koetsier2010} which would require lower temperatures.  Finally, the achievement of cold spin-imbalanced clouds in an optical lattice suggests a new route for local entropy reduction using adiabatic demagnetization cooling, a technique previously demonstrated in bosonic lattice experiments \cite{Medley2011}. 
\raggedbottom
\paragraph{Acknowledgements}
This work was supported by the NSF (grant no. DMR-1607277), the David and Lucile Packard Foundation (grant no. 2016-65128), and the AFOSR Young Investigator Research Program (grant no. FA9550-16-1-0269). W.S.B. was supported by an Alfred P. Sloan Foundation fellowship. P.T.B. was supported by the DoD through the NDSEG Fellowship Program. E.K. was supported by the NSF (grant no. DMR-1609560). T.P. was supported by CNPq, FAPERJ, and INCT on Quantum Information. N.T. acknowledges funding from the NSF (grant no. DMR-1309461). We thank M. Rigol for providing exact diagonalization results for the imbalanced system at $U=0$ to benchmark the NLCE. 
\paragraph{Author Contributions}
P.T.B., D.M., E.G.-S., P.S., and S.K. performed the experiment and data analysis. E.K. performed the NLCE calculations. T.P. and N.T. performed the DQMC calculations. W.S.B. and D.A.H. provided scientific guidance in experimental and theoretical questions. All authors contributed to the interpretation of the data and the writing of the manuscript.

%


\pagebreak
\clearpage
\setcounter{equation}{0}
\setcounter{figure}{0}

\renewcommand{\theparagraph}{\bf}
\renewcommand{\thefigure}{S\arabic{figure}}
\renewcommand{\theequation}{S\arabic{equation}}

\onecolumngrid
\flushbottom

\section{\Large Methods}

{\bf Preparation of a spin-imbalanced gas in an optical lattice} We realize the Fermi-Hubbard model using a degenerate mixture of two Zeeman states ($\ket{1}=\ket{\uparrow}$ 	and $\ket{3}=\ket{\downarrow}$, numbered up from lowest energy) in the ground state hyperfine manifold of $^6$Li placed in an optical lattice. The global spin imbalance $P=\frac{N_1-N_3}{N_1+N_3}$ can be varied continuously from 0 to $\approx0.9$ while the temperature $T$ remains relatively unaffected. 

To create the sample we load a magneto-optical trap (MOT) from a Zeeman slower, then use a compressed MOT stage before loading into a $\approx$ \SI{1}{\milli \kelvin} deep optical trap and evaporating near the \SI{690}{G} Feshbach resonance. We stop the evaporation before Feshbach molecules form and transfer the atoms to a highly anisotropic `light sheet' trap with aspect ratio $\omega_x:\omega_y:\omega_z=1:2:10$ where it undergoes further evaporation near \SI{300}{G} where $a_s = -890 a_0$. Next we transfer to the final trap geometry where a \SI{1070}{\nm} beam provides radial confinement and a \SI{532}{\nano\meter} accordion lattice with trapping frequency $\omega_z = 2 \pi \times$ \SI{19.9(3)}{\kilo\hertz} provides axial confinement (for further details see \cite{SuppMitra2016}). The spin populations are imbalanced by evaporating the mixture in a magnetic gradient of up to \SI{40}{G \per \centi \meter} along the same direction as the magnetic bias field, which we set in the range 75-\SI{500}{G} depending on the targeted imbalance. We then tune the bias field to set the scattering length and load into a 2D square lattice with a \SI{50}{\milli\second} long ramp to a depth of $10.5(3) E_r$.

\begin{figure*}[h]
\centering
\includegraphics[width=0.35\textwidth]{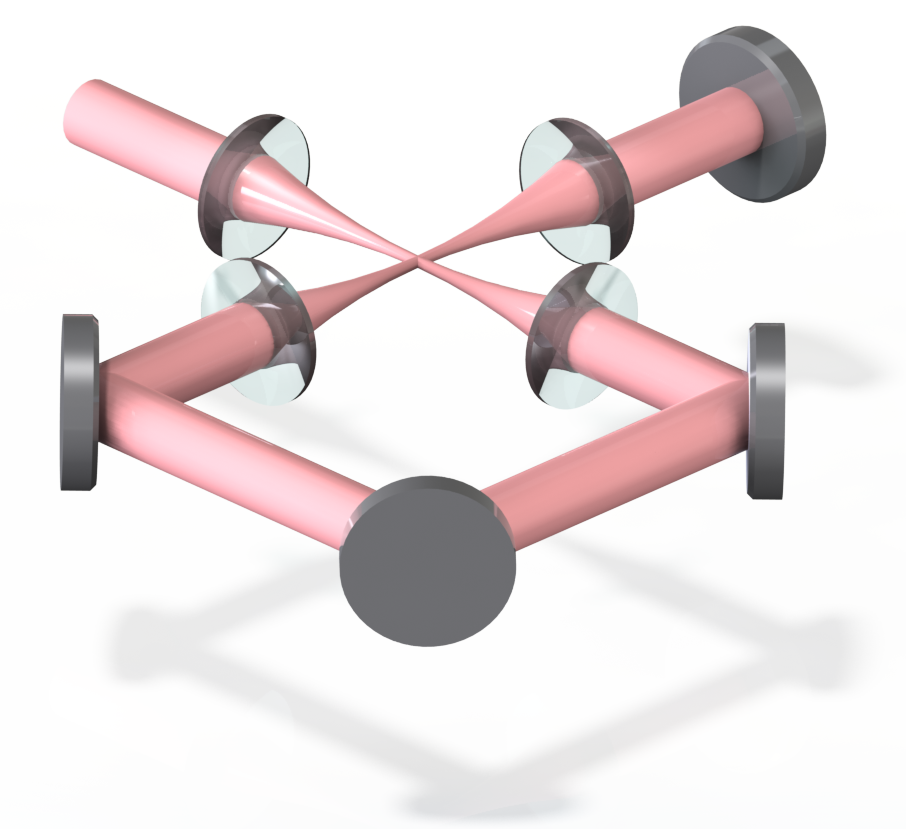}
\caption{{\bfseries Lattice geometry} Lattice beams are shown in pink, the mirrors in gray and the lenses in blue. All four lattice beams crossing in the center interfere in this configuration.}
\label{fig:lattice}
\end{figure*}

{\bf Details of the 2D lattice} We use a non-separable 2D lattice with \SI{752}{\nano\meter} spacing formed by four interfering passes of a single vertically polarized beam. The lattice geometry is shown in Fig.~\ref{fig:lattice} and described in \cite{SuppSebby-Strabley2006}.  Compared to the commonly used lattice configuration created by two non-interfering orthogonal retro-reflected beams, this lattice has a spacing a factor of $\sqrt{2}$ larger and a much larger depth because of the 4-fold interference, features that facilitate quantum gas microscopy.

Here we have to take into account that the two incoming lattice beams are not exactly at a \SI{90}{\degree} angle. To compensate the resulting tunneling asymmetry in the lattice and retain a symmetric lattice we attenuate the retro-reflected laser power. The potential at the atoms is given by

\begin{eqnarray*}
V(x,y) = V_0 \left(1- \frac{1 + r^2 + 2 r \cos\left(2 k x \cos(\theta/2)\right)}{1+2r+r^2} \cdot \frac{1+\cos\left(2 k y \sin(\theta/2)\right)}{2} \right)
\end{eqnarray*}
where we align the coordinate axes with the principle axes of the lattice. The lattice beams travel in the $\mathbf{x}+\mathbf{y}$ and $\mathbf{x}-\mathbf{y}$ directions. Here, $V_0$ is the full lattice depth, $\theta$ the angle between the first two incoming lattice beams, $k = 2\pi/\lambda$ and $r$ the electric field amplitude attenuation factor of the retro-reflected beam. For $\theta \approx \SI{90}{\degree}$ the lattice constant is $a_\text{lat} \approx \frac{\lambda}{\sqrt{2}} = \SI{752}{nm}$. The lattice beam waist is \SI{70}{\micro \meter}.

{\bf Calibration of Hubbard parameters} We extract our lattice parameters from atom resolved images, finding that the lattice axes are at an angle of $\SI{90.02(2)}{\degree}$ and the lattice constants are $a_x = \SI{762}{\nano\meter}$ and $a_y = \SI{741}{\nano\meter}$ from which we extract the intersection angle of the lattice beams, $\theta = 2\arctan\left(a_x/a_y\right) = \SI{91.63(2)}{\degree}$. 

We experimentally compensate for the tunneling asymmetry by finding the retro-reflection power that gives symmetric tunneling by making the singles-density correlators equal along both lattice axes. We calibrate our lattice depth by measuring the frequencies of the three $d$ bands in a deep lattice using lattice amplitude modulation, and compare these with the 2D band structure calculation, including the measured lattice angle and the retro-reflection attenuation factor $r$ as a fit parameter. At the best fit $r=0.54$, the inferred depth of the lattice at which our measurements are performed is \SI{10.5(3)}~$E_r$, where $E_r = \SI{14.66}{\kilo\hertz}$. From that we obtain tight-binding tunneling values $t_x = 442$ Hz, $t_y = 462$ Hz ($t_x/t_y = 0.96$), confirming our experimental procedure for making the tunneling symmetric. The reduction of the lattice depth across the cloud due to the gaussian profile of the lattice beams leads to an increase in the tunneling by \SI{10}{\%} at the edge of the cloud compared to the central value. We also have a non-zero but negligible diagonal tunneling $t_d = \SI{12}{Hz} = 0.03 t_x$, due to the non-separability of the lattice

We measure the interaction energy $U$ using radio frequency spectroscopy. We transfer atoms from state $\ket{1}$ to $\ket{2}$ and resolve the frequency shift between singly and doubly occupied sites. We determine $U_{13} = \delta U \frac{a_{13}}{a_{13}-a_{23}}$ taking into account a small correction due to weak final state interactions. The experimentally measured value agrees with the value determined from band structure calculations including higher band corrections \cite{SuppIdziaszek2005} to within 10\%. 	

   \begin{figure*}[h]
    \centering
   \includegraphics[width=0.5\textwidth]{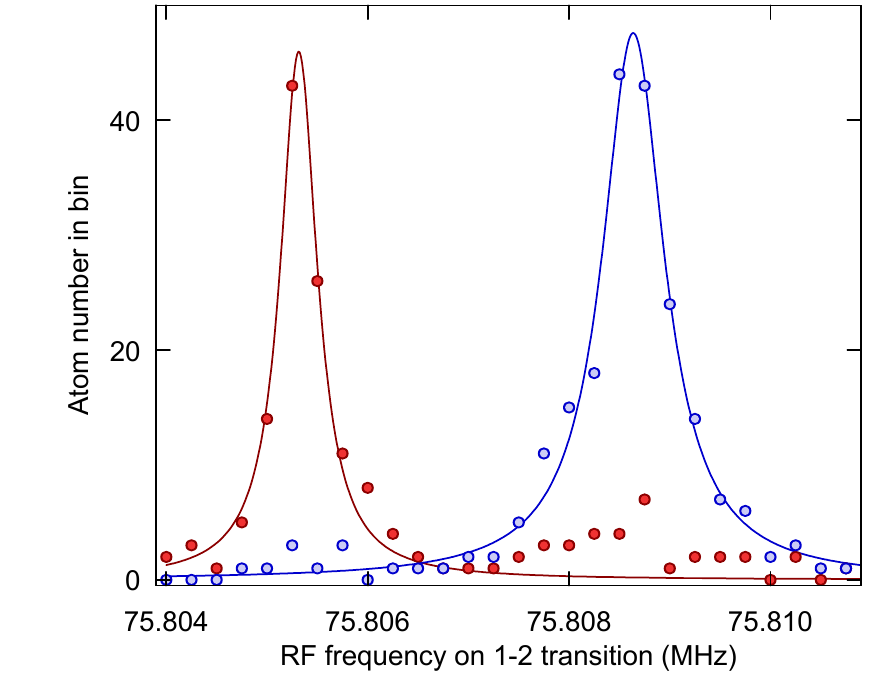}
    \caption{ {\bfseries RF spectroscopy for $U$-calibration.} Spectroscopy signal from a band insulator at lattice depth of $10.5 E_r$ and scattering length $448 a_0$. Shown is the transferred atom number in the center of the cloud (red), where we had mainly doubly occupied sites and in the surrounding Mott insulator region (blue) where we measure a clean bare resonance signal. To determine the interaction energy the final state interaction has to be taken into account (see text).\label{fig:interaction}}
   
  \end{figure*}

{\bf Measurement of spin correlations} After the imbalanced gas is loaded adiabatically into the optical lattice, we freeze tunneling dynamics by increasing the lattice depth within \SI{100}{\micro \second} to 55 $E_r$. After that, we convert the $\ket{1} - \ket{3}$ mixture to a $\ket{1} - \ket{2}$ mixture by driving a radio-frequency Landau-Zener transition between states $\ket{3}$ and $\ket{2}$ in \SI{10}{\milli\second} with efficiency of $0.99(1)$. This step is necessary because an RF transition between state $\ket{1}$ and $\ket{3}$ is forbidden, so we cannot perform pseudospin rotations with the $\ket{1}-\ket{3}$ mixture. Next we hide all doublons in our sample by converting them to molecules, employing a ramp across a narrow Feshbach resonance near \SI{543}{G}. Then we either drive a $\pi/2$ spin rotation to measure correlations for the spin component perpendicular to the effective magnetic field, or omit this to measure correlations for the spin component parallel to the field.

We image the final atom distribution by increasing the lattice depth to 2500 $E_r$ within \SI{250}{\micro \second}, ramping up the light sheet to provide axial confinement, and then collecting fluorescence photons during Raman cooling. Raman imaging is not spin sensitive, so to measure spin correlations we remove one spin state from the trap before imaging. We take three sets of images, one with $|1\rangle$ atoms removed, one with $|2\rangle$ atoms removed, and one with no removal pulse. 

{\bf Spin removal} For removing atoms in a particular spin state, we illuminate the cloud with a \SI{30}{\micro\second} resonant pulse. The length of this pulse was determined by observing atom loss for different pulse lengths. The atom loss curve exhibits two well-separated time scales, which we interpret as resonant heating and off-resonant heating of the two spin states, and fit with the sum of two decaying exponentials \cite{SuppParsons2016}. We find time constants \SI{4.5(2)}{\micro\second} and \SI{1.7(5)}{\milli\second}, from which we estimate that $0.2\%$ of resonantly blown atoms remain, and up to $2\%$ of the off-resonant state atoms are ejected.

The resonant pulse also pumps some of state $\left| 1 \right>$ ($\left| 2 \right>$) atoms into state $\left| 5 \right>$ ($\left| 4 \right>$). We take advantage of the cycling state $\left| 3 \right> = \left|m_I = -1, m_J = -1/2 \right>$ to calibrate these values. To measure the probability of pumping $\left| 1 \right>$ into $\left| 5 \right>$ we prepare a $\left| 1 \right>-\left| 3 \right>$ Mott insulator and compare the results blowing both states with blowing only state $\left| 3 \right>$. We find that we pump \SI{0.8(1)}{\%} of the atoms from $\left| 1 \right>$ to $\left| 5 \right>$. We separately measured the probability of pumping $\left| 2 \right>$ into state $\left| 4 \right>$ to be \SI{1.2(1)}{\%} by preparing a $\left| 1 \right>-\left| 2 \right>$ Mott insulator and comparing blowing both states with blowing $\left| 2 \right>$ only. 

{\bf Handling doublons} We observe that resonantly blowing one of the spin states does not lead to loss of a doublon, but ejects only a single atom. The other atom, which has a different spin state, remains behind. This is in contrast to the behavior observed during Raman imaging, where doublons are lost through light-assisted collisions. As the spin correlator cannot be easily extracted with the extra atoms from the doublon blowing, we hide doublons by ramping the magnetic field over the narrow $\ket{1}-\ket{2}$ Feshbach resonance near \SI{543}{G} thereby converting doublons to Feshbach molecules before resonantly blowing.

To estimate the fidelity of doublon hiding, we first prepare a band insulator. We take images with neither doublon hiding nor blowing to determine the number of singles in the band insulator. We then compare this to the number of singles we observe after performing doublon hiding and removing one of the spin states with a blowing pulse, which leaves behind single atoms on sites where the doublon hiding failed. We find a doublon hiding fidelity of $90(3)\%$.

{\bf Raman imaging} We use a Raman cooling scheme for lithium similar to the ones demonstrated before \cite{SuppOmran2015,SuppParsons2015} but with slight differences. We use a lin-perp-lin configuration for the Raman beams in a retro-reflected configuration. The Raman axis is along the diagonal of the square lattice and tilted about $\SI{10}{\degree}$ out of plane. The optical pumping beam is circularly polarized. We use a magnetic offset field of about \SI{200}{mG} which is in the plane of the lattice and perpendicular to the Raman axis. The optical pumping beam is tuned \SI{5}{\giga\hertz} to the red of the $D_1$ line. We typically operate with \SI{50}{\micro\watt} focused to a waist of about \SI{1}{\milli\meter} for the pump beam and \SI{3}{\milli\watt} focused to a waist of \SI{75}{\micro\meter} for each Raman beam. We estimate our Rabi frequency  to be $\Omega = (2 \pi)$ \SI{180}{\kilo\hertz} by measuring power broadened Raman spectra for different powers.

   \begin{figure*}[ht]
   \centering
   \includegraphics[width=0.9\textwidth]{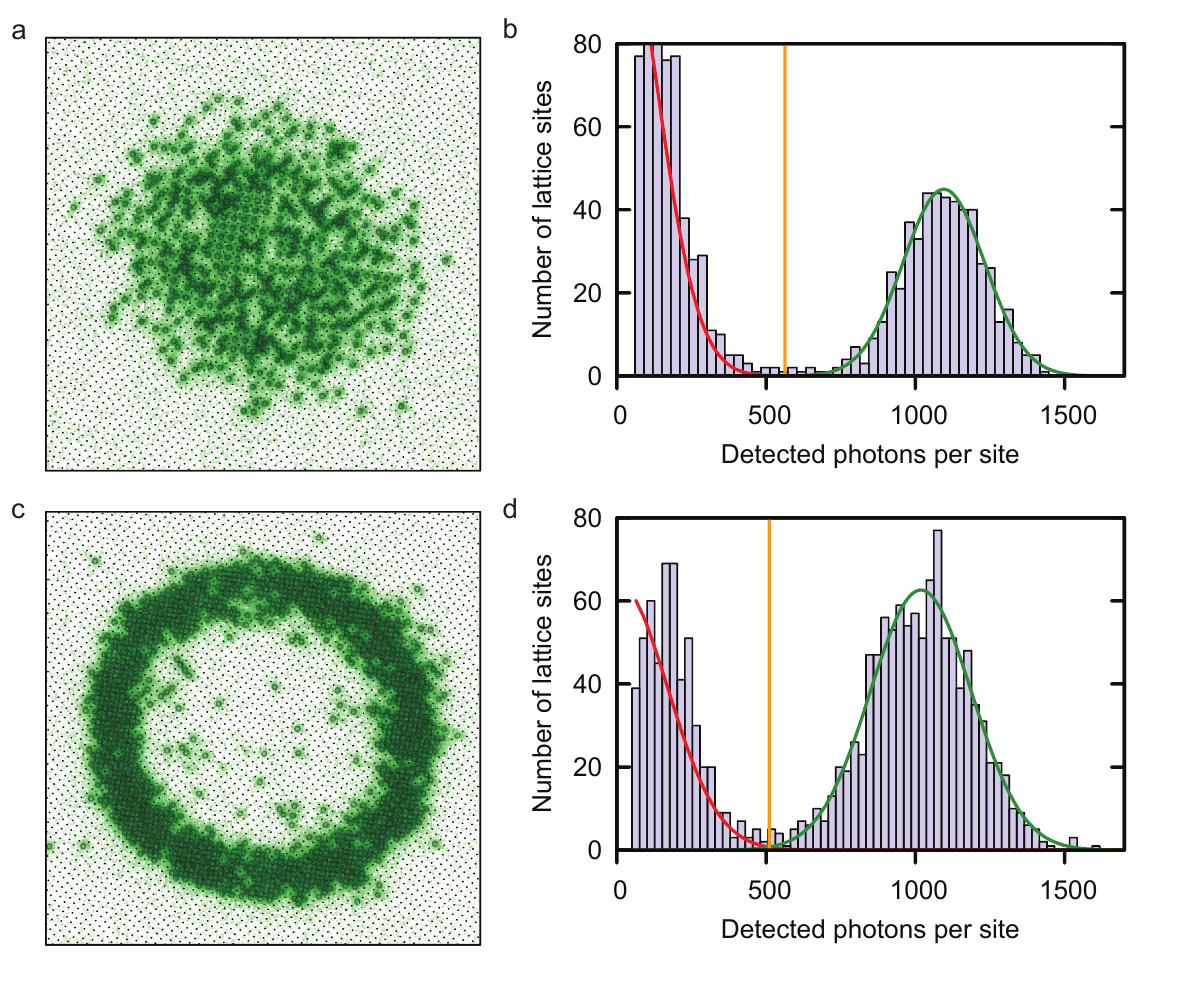}
    \caption{{\bfseries Reconstruction visualization} \textbf{a}, Fluorescence image from Fig. 1c of the main text with lattice sites overlaid, showing occupied sites (circles) and unoccupied sites (points). Field of view is \SI{46}{\micro\meter}. \textbf{b}, Histogram of detected photons on each site for panel a. We identify the lower peak with unoccupied sites and the upper with occupied sites. By fitting gaussians to these peaks, we identify a threshold value (orange line). Any site with more counts than the threshold we count as an atom. \textbf{c}, Fluorescence image showing a band insulator in the center of the cloud, surrounded by a Mott insulator region. \textbf{d}, Histogram of detected photons for panel c.}
    \label{fig:reconstr}
  \end{figure*}

We perform Raman imaging for \SI{1200}{\milli \second} and collect approximately $1000$ photons per atom using a Special Optics custom objective with a working distance of $24.7$\,mm. An achromatic doublet with $f = \SI{750}{mm}$ gives a magnification of $\approx 30$, which we verified using Kapitza-Dirac scattering of a molecular BEC. Our objective is designed to be diffraction limited for \SI{671}{\nano\meter} and corrected for our \SI{5}{\milli\meter} thick fused silica vacuum window. We image the atoms with a Zyla 4.2 CMOS camera (Andor Technology) with quantum efficiency of $\approx 75$\,\% near \SI{671}{\nano\meter}. 

{\bf Image analysis} We reconstruct the atom distribution in the lattice following the image reconstruction procedure described in \cite{SuppSherson2010} (Fig. \ref{fig:reconstr}). The reconstruction allows us to reduce the data from the fluorescence pictures to binary matrices that are the basis for all further data processing. Our measured point spread function (PSF) has a full-width half-max of \SI{900(20)}{\nano\meter}, slightly larger than expected for our numerical aperture of 0.5.

We estimate fidelity errors due to Raman imaging imperfections by taking 40 consecutive images with our standard imaging time of \SI{1200}{\milli\second} of the same atom cloud and determine the shot-to-shot differences. This leads to a hopping rate during one picture of \SI{0.4(2)}{\%} and a loss rate of \SI{1.6(3)}{\%}. In addition, while holding the atoms in a deep lattice for spin manipulations and doublon hiding, we lose \SI{2(1)}{\%} of the atoms, leading to a net detection efficiency of $\approx \SI{96}{\%}$.

{\bf Extracting density and spin correlations} We are able to measure several different kinds of correlators with the techniques described previously. Recall that we define the singles density of spin $\sigma$ on lattice site $\mathbf{i} = (i_x,i_y)$ by $n^s_{\mathbf{i},\sigma} = n_{\mathbf{i},\sigma} - n_{\mathbf{i},\uparrow} n_{\mathbf{i},\downarrow}$. We also define $n^s_{\mathbf{i}} = \sum_\sigma n^s_{\mathbf{i},\sigma}$.  

The first correlator we measure is the singles density correlator. This is the natural quantity we measure with the microscope without any blowing pulses or doublon hiding. In this case, the doublons are removed by light-assisted collisions during the imaging. The correlator is defined by,
\[M(\mathbf{d}) = \left<n^s_\mathbf{i} n^s_{\mathbf{i}+\mathbf{d}} \right>_c,\]
where $\left<AB\right>_c = \left<AB\right> - \left<A\right>\left<B\right>$.
If we insert a resonant blowing pulse to eject one spin state and additionally perform doublon hiding, we measure singles density correlations of one spin state,
\[B_\sigma(\mathbf{d}) = \left<n^s_{\mathbf{i},\sigma} n^s_{\mathbf{i}+\mathbf{d},\sigma} \right>_c.\]
This correlator is closely related to the spin correlator, but with additional contributions due to the presence of holes and doublons.

To define the spin correlator, we first write the $z$-component of the spin operator in terms of density operators, $S^z_\mathbf{i} = \frac{1}{2}(n_{\mathbf{i},\uparrow}-n_{\mathbf{i},\downarrow})$. Then, we have
\begin{eqnarray*}
C^z(\mathbf{d}) &=& 4\left<S^z_\mathbf{i} S^z_{\mathbf{i}+\mathbf{d}} \right>_c\\
&=& \left<(n_{\mathbf{i}\uparrow}-n_{\mathbf{i}\downarrow})(n_{\mathbf{i}+\mathbf{d},\uparrow}-n_{\mathbf{i}+\mathbf{d},\downarrow}) \right>_c \\
&=& 2 \left[B_\uparrow(\mathbf{d}) + B_\downarrow(\mathbf{d})\right] - M(\mathbf{d}).
\end{eqnarray*}


We define $C^x$ and $C^y$ analogously. In the experiment we cannot distinguish between $S^x$ and $S^y$ spin correlations and effectively measure $C^\perp = \frac{1}{2} \left(C^x + C^y \right)$. The Fermi-Hubbard model with an applied field along the $z$-direction retains in-plane rotational symmetry, so $C^\perp = C^x = C^y$. 

{\bf Systematic errors in measuring correlations} Atoms that are lost from the trap after tunneling dynamics are frozen can reduce our measured correlations. We estimate this effect by introducing the probability of an atom being lost before imaging, $\epsilon_l$, and the probability $\epsilon_r$ that we image a single when it is not present on a site. $\epsilon_r$ is a result of both hopping and recapture of atoms during imaging. We find
\begin{eqnarray*}
\tilde{n}^s_\mathbf{i} &=& n^s_\mathbf{i} (1-\epsilon_l) + (1 - n^s_\mathbf{i}) \epsilon_r \\
\left<n^s_\mathbf{i} n^s_{\mathbf{i}+\mathbf{d}} \right>_c - \left< \tilde{n}^s_\mathbf{i} \tilde{n}^s_{\mathbf{i}+\mathbf{d}}\right>_c &=& 2\left(\epsilon_l+\epsilon_r\right)\left<n^s_\mathbf{i} n^s_{\mathbf{i}+\mathbf{d}} \right>_c, \\
\end{eqnarray*}
to first order in $\epsilon$. Using our previous estimates for imaging loss rate, hopping rate, and loss during hold times, leads to an expected reduction of $0.92(2)$ in all measured correlators.

Imperfections in the spin imaging process can also effect correlators. Our blowing pulse off-resonantly ejects an atom in the spin state of interest with probability $\epsilon_{\sigma}$ and fails to eject an atom in the other spin state with probability $\epsilon_{f\sigma}$, leading to
\[\tilde{n}^s_{\mathbf{i},\sigma} = n^s_{\mathbf{i},\sigma} (1 -\epsilon_{\sigma}) + n^s_{\mathbf{i},-\sigma} \epsilon_{f \sigma}.\]
The errors considered here are due to the resonant blowing pulses and do not effect the value of $M(\mathbf{d})$. However, they do effect the $C^z(\mathbf{d})$ through the $B_\sigma(\mathbf{d})$. The reduction in these correlators is
\begin{eqnarray*}
B_\sigma(\mathbf{d}) - \tilde{B}_\sigma(\mathbf{d}) &=& \left(2\epsilon_\sigma \right) B_\sigma(\mathbf{d}) - \left(\epsilon_{f,\sigma}\right) \left[\left< n^s_{\mathbf{i},\sigma} n^s_{\mathbf{i}+\mathbf{d},-\sigma}\right>_c + \left<n^s_{\mathbf{i},-\sigma} n^s_{\mathbf{i}+\mathbf{d},\sigma} \right>_c \right] \\
C^z(\mathbf{d}) - \tilde{C}^z(\mathbf{d}) &=& \left(\epsilon_{f,\uparrow} + \epsilon_{f,\downarrow}\right) C^z(\mathbf{d}) + \sum_\sigma \left(\epsilon_{f\uparrow} + \epsilon_{f\downarrow} - 2 \epsilon_\sigma \right) B_\sigma(\mathbf{d}),
\end{eqnarray*}
where we have used the identity $C^z(\mathbf{d}) - B_\uparrow(\mathbf{d})-B_\downarrow(\mathbf{d}) = \left< n^s_{\mathbf{i},\uparrow} n^s_{\mathbf{i}+\mathbf{d},\downarrow}\right>_c + \left<n^s_{\mathbf{i},\downarrow} n^s_{\mathbf{i}+\mathbf{d},\uparrow} \right>_c$. For the spin correlator, the magnitude of the first term is at most $1 \times 10^{-3}$, while the second two are at most $5 \times 10^{-3}$.

Imperfect doublon hiding also effects our correlators. If we fail to hide a doublon with probability $\epsilon_d$, we find
\begin{eqnarray*}
\tilde{n}^s_\mathbf{i} &=& n^s_\mathbf{i} + \epsilon_d d_\mathbf{i}\\
\left<n^s_\mathbf{i} n^s_{\mathbf{i}+\mathbf{d}} \right>_c - \left< \tilde{n}^s_\mathbf{i} \tilde{n}^s_{\mathbf{i}+\mathbf{d}} \right>_c &=& -\left(\epsilon_d \right) \left(\left< n^s_\mathbf{i} d_{\mathbf{i}+\mathbf{d}} \right>_c + \left<d_\mathbf{i} n^s_{\mathbf{i}+\mathbf{d}} \right>_c\right).
\end{eqnarray*}
NLCE predicts the correlator on the right hand side of this equation is at most $3 \times 10^{-3}$ (see supplement to \cite{SuppCheuk2016}). This effect has magnitude at most $3 \times 10^{-4}$.

{\bf The Numerical Linked Cluster Expansion} The numerical linked-cluster expansions (NLCEs)~\cite{SuppRigol2006,SuppTang2013} combine high-temperature series expansions (HTSEs) with exact diagonalization techniques to calculate exact finite-temperature properties of quantum lattice models in an extended temperature range in the thermodynamic limit. The formalism can be summarized in the following equation
\begin{equation}
P= \sum_c W_P(c),
\label{eq:nlce}
\end{equation}
where $P$, an extensive property per lattice site in the thermodynamic limit, is expressed in terms of contributions [or weights, $W_P(c)$] of all the clusters $c$ that can be embedded in the lattice and are not related by any translational symmetry. The smallest cluster is a single site, the next smallest is a nearest-neighbor bond, and so on. If the Hamiltonian is also invariant under point group symmetries of the underlying lattice, the number of terms in Eq.~\ref{eq:nlce} can be reduced  by considering only clusters that are topologically distinct and multiplying their weights by a multiplicity factor $M(c)$, which represents the number of ways per site cluster $c$ can be embedded on the lattice. The weights $W_P(c)$ are in turn calculated using the inclusion-exclusion principle; writing Eq.~\ref{eq:nlce} for the finite cluster $c$, we can express $W_P(c)$ as 
\begin{eqnarray}
W_P(c)&=& p(c) - \sum_{s\subset c} W_P(s),
\end{eqnarray}
where $p(c)$ is the extensive property of interest on the finite cluster $c$, and $s$ are all the clusters that can be embedded in $c$ (sub-clusters of $c$).

Unlike in the HTSEs, where $p(c)$ are expressed as perturbative expansions in terms of powers of the inverse temperature, in the NLCEs, we obtain $p(c)$ exactly (to all powers in the inverse temperature) using full diagonalization, leading to a better convergence.

We use the site expansion NLCE in which the results in the $n$th order contain contributions from all clusters with $n$ sites or less. We carry out the expansion to the 9th order. Only in the last order, and for each of the two values of $U/t$ (8.0, and 14.0), we diagonalize the Hubbard Hamiltonian in the presence of a magnetic field for 112 topologically distinct 9-site clusters  that have no particular symmetry. We choose a fine grid for the temperature and work in the grand canonical ensemble; for $U/t=8.0$ at half filling, we fix the chemical potential and choose a fine grid for the magnetic field. Similarly for $U/t=14$ away from half filling, we fix the magnetic field and choose a fine grid for the chemical potential. These allow us to have a high resolutions for the polarization and the density, respectively~\cite{SuppKhatami2011}.

The converged results are valid in the thermodynamic limit and contain no systematic or statistical errors. However, due to the finite number of terms in our series the convergence is lost below a temperature that is dependent on $U/t$, the polarization and the density. To extend the convergence to lower temperatures, we take advantage of numerical re-summation techniques. To minimize the possibility of introducing systematic errors, we use two vastly different algorithms, namely, Wynn resummation with four cycles of improvement and the Euler resummation for the last six terms in the series~\cite{SuppRigol2007,SuppTang2013}, and present results when they agree within a few percent.

\begin{figure*}[h]
   \centering
   \includegraphics[width=\textwidth]{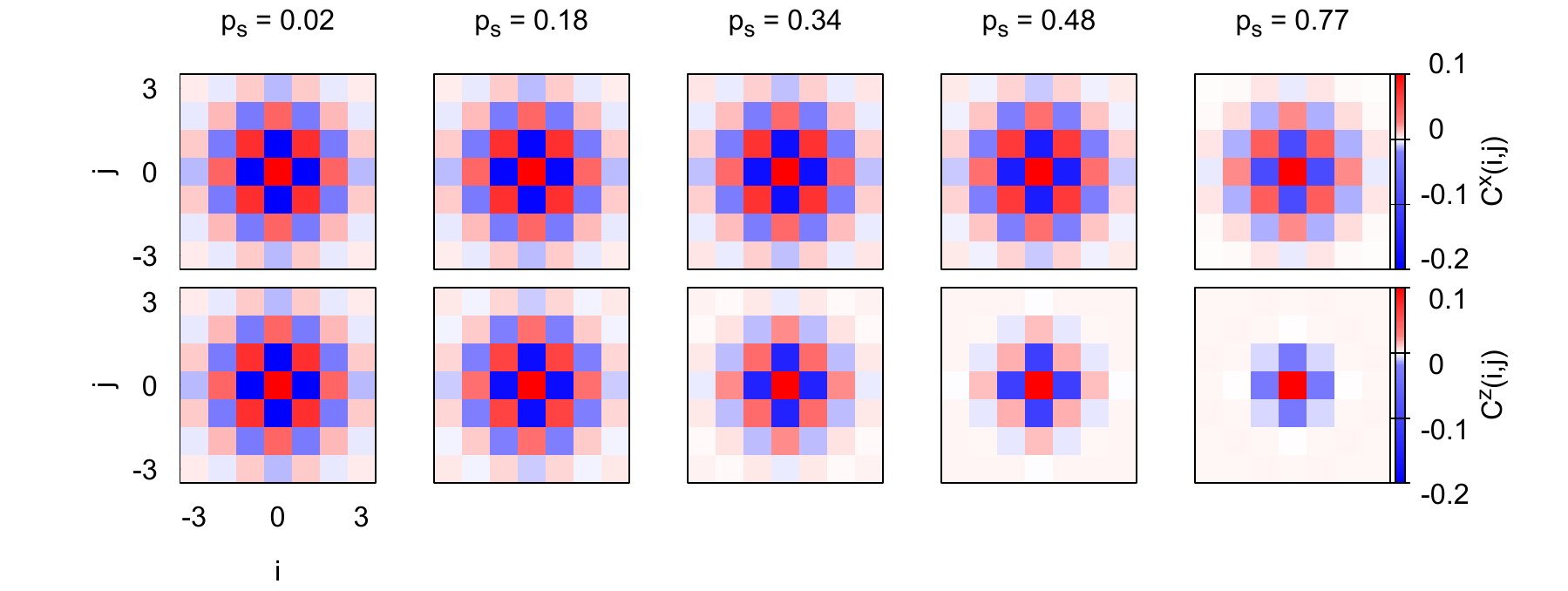}
    \caption{{\bfseries Spin correlators vs. singles polarization} NLCE results for $T/t=0.415$. Top row shows $S^x$ correlators, $C^x$, and bottom row shows $S^z$ correlators, $C^z$.}
    \label{fig:spincorr}
  \end{figure*}

In the NLCE, one can in principle compute correlation functions at distances as far as the largest clusters in the series extend. However, as this distance increases, a smaller number of clusters contribute to the series, and so, one should expect the lowest convergence temperature to increase with distance. In Fig. \ref{fig:spincorr}, we show spin correlations from the NLCE at $U/t=8.0$ for distances as large as $(3,3)$. We find a very good qualitative agreement with the experimental results shown in Fig. 3 of the main text. The NLCE results are at a fixed $T/t=0.415$, so we do not consider the temperature variations that may exist from one polarization to the next in the experimental data.

{\bf Determinantal Quantum Monte Carlo} The determinantal or auxilliary field quantum Monte Carlo (QMC) for interacting fermions is an unbiased algorithm that provides statistically ``exact" answers for the energy and correlation functions as a function of temperature. We start with the Hubbard model in terms of fermion operators $H=T+V$ where the single particle part of the Hamiltonian is

\begin{equation}
T=-t\sum_{\langle i,j\rangle\sigma}[c_{i\sigma}^\dagger c_{j \sigma} + h.c.] - \mu \sum_{i} (n_{i\uparrow }+ n_{i\downarrow }) - h  \sum_{i} (n_{i\uparrow }- n_{i\downarrow })
\end{equation}
and the repulsive ($U>0$) interacting part 
\begin{equation}
V=U\sum_i ( n_{i\uparrow}-1/2  ) ( n_{i\downarrow}-1/2  ).
\end{equation}

Here, $t$ is the hopping amplitude between sites, $\mu$ is the chemical potential that controls the density of fermions and $h$ is the Zeeman field that controls the imbalance between majority up and minority down spin fermions. The experiments are performed in the canonical ensemble for a fixed total number of fermions $N=N_\uparrow+N_\downarrow$ and a fixed magnetization $M=N_\uparrow-N_\downarrow$ whereas the simulations are done in the grand canonical ensemble for a fixed $\mu$ and $h$.

For a bipartite lattice, the repulsive Hubbard model maps onto an attractive Hubbard model through a particle-hole transformation on the down spins and an additional $\pi$ phase shift on the B sublattice given by $c_{A\downarrow}\rightarrow c^\dagger_{A\downarrow}$, $c_{B\downarrow}\rightarrow -c^\dagger_{B\downarrow}$. With this mapping the chemical potential and Zeeman field get interchanged $\mu\leftrightarrow h$ and $h\leftrightarrow \mu$ in the repulsive and attractive Hubbard models. We will return to this mapping when we discuss the nature of the sign problem toward the end of the section.
 
With the Hamiltonian defined above, we next divide up the imaginary time interval $[0,\beta=1/(k_B T)]$ into $M$ segments of width $\delta\tau=\beta/M$. A Trotter break up of the two non-commuting operators $T$ and $V$ can now be done on each time-slice with errors on the order of $t U (\delta \tau)^2$, which gives $e^{-\beta H} = [e^{-\delta \tau (T+V)}]^M\approx 
[e^{-\delta \tau T} e^{-\delta \tau V}]^M $. By introducing auxiliary fields at each space-time point $\vec S(i,\tau)$ in a path integral representation of the partition function in imaginary time, typically we use Ising auxiliary fields $S= \pm 1$ at each space-time point, the quartic interaction terms can be factored into quadratic forms, given by
\begin{eqnarray}
e^{-\delta \tau U ( n_{i\uparrow}-1/2  ) ( n_{i\downarrow}-1/2  )} &=& {\frac{1}{2}} e^{- { \delta\tau U (n_{i\uparrow}+ n_{i\downarrow}) /2}}\sum_{S(i,\tau)=\pm 1} e^{- \lambda S(i,\tau) (n_{i\uparrow}- n_{i\downarrow})}\\ \nonumber
&=&  {\frac{1}{2}} \sum_{S(i,\tau)=\pm 1} \prod_{\sigma=\pm 1} e^{- ( \sigma \lambda S(i,\tau) + U \delta \tau/2)n_{i\sigma}}
\end{eqnarray}
with $\lambda$ determined by $\rm cosh( \lambda) = \exp(\delta \tau U/2)$ and  $\sigma=\pm 1$.

The path integral of the interacting problem is thus converted to an effectively non-interacting action that describes the motion of a fermion propagating through fluctuating auxiliary field configurations, given by:
\begin{eqnarray}
-t\sum_{\langle i,j\rangle\sigma}[c_{i\sigma}^\dagger h_\sigma (\bar S(\tau))c_{j \sigma} ] &=& -t\sum_{\langle i,j\rangle\sigma} [c_{i\sigma}^\dagger c_{j \sigma}  + h.c.] \\ \nonumber
&+& \sum_{i \sigma}  [-(\mu +U/2+\sigma h) +\sigma \lambda S(i,\tau)] n_{i \sigma}.
\end{eqnarray} 
The fermions can now be integrated out yielding a partition function 
\begin{equation}
Z=\sum_{\lbrace S \rbrace}{\rm Det} M_\uparrow (S)  { \rm Det}M_\downarrow (S)
\end{equation}
as a sum of determinants over auxiliary field configurations that are sampled using Monte Carlo methods, where $M_\sigma (S) = [I+\prod_\tau \exp(-\delta\tau h_\sigma(S(\tau))]$. The negative sign or complex phase of this determinant is the source of the so-called ``sign-problem" in QMC simulations that can limit the ability to go to low temperatures away from half filling. 

\begin{figure}[!tb] 
 \includegraphics[width=0.65\columnwidth]{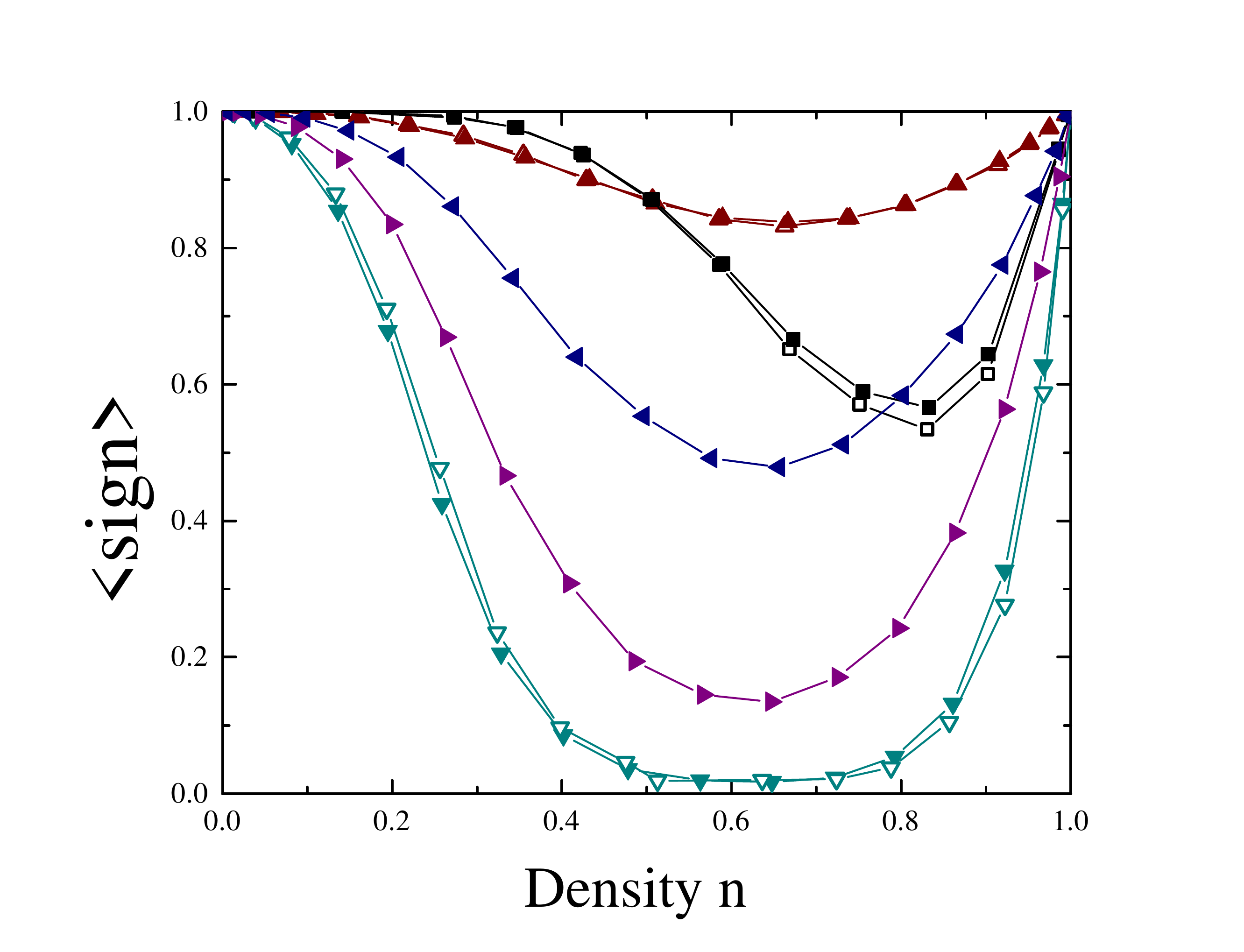}
 \vspace{-0.1cm}
 \caption{Average sign as a function of density for  $T/t=0.36$ (teal and black), $T/t=0.42$ (purple), $T/t=0.50$ (blue), and $T/t=0.63$ (brown), open symbols for balanced populations and closed symbols for $h/t=0.20$, triangles for $U/t=15$ and squares for $U/t=8$.}
 \label{fig:sign}
\end{figure}

In Fig. \ref{fig:sign} we show the average sign as a function of density at various temperatures for zero field and a finite field. The important points to note are: (i) The magnetic field does not generate a sign problem since it essentially maps to $\mu$ in the attractive Hubbard model for which the up and down determinants have the same sign, hence the product is always positive, for any density. (ii) At half filling the average sign is unity, hence no sign problem, for any temperature and field. (iii) For $U/t=8$ (Fig.~\ref{fig:sign}, black squares) where most experimental data were taken, the sign problem is not severe and does not affect the quality of the data at the lowest temperatures considered. (iii) Away from half filling the sign is most severe  around $0.5 \lesssim  n \lesssim 0.8$ filling for $U/t=15$; the average sign nose dives with decreasing temperature. Within this density region, for $T/t  \ge 0.42$ increased statistics is used to improve the quality of the data and we were able compare our QMC with the experimental data for the density and the polarization. For even smaller temperatures (Fig.~\ref{fig:sign}, down triangles)  we have not used QMC data to compare with experiments.
The error bars reported in the figures are statistical. The QMC simulations were performed for $8\times 8$ lattices with $ t  \delta\tau = 0.05$ for $U/t=8$ and  $ t  \delta\tau = 0.02$ for $U/t=15$. Each data point was obtained from up
to 60 independent runs, each with 4000 sweeps through the lattice.

{\bf Sign change of spin $z$-correlation for large polarizations} To understand why $C^z(1,1)$ becomes negative near $p_s = 0.6$ at half filling, it is instructive to consider what happens in the Heisenberg model with a nearly fully polarized gas. We can regard the fully polarized gas as a vacuum state and the minority spins as a dilute gas of magnons \cite{SuppBatyev1984,SuppZapf2014}. These magnons are bosons and can form a Bose Einstein Condensate (BEC) at low temperatures. We will find that the BEC off-diagonal order is associated with the spin correlations perpendicular to the magnetization, and the density correlation of the magnons are associated with the spin correlations parallel to the magnetization. 

To make this argument more concrete, we rewrite the Heisenberg Hamiltonian using operators defined by $\beta^{\dag}_\mathbf{i} = S^+_\mathbf{i}$, $\beta_\mathbf{i} = S^-_\mathbf{i}$ and  $S^z_\mathbf{i} = \beta^{\dag}_\mathbf{i}\beta_\mathbf{i} - 1/2$,
\begin{equation}\mathcal{H} = \frac{J}{2} \sum_{\left<\mathbf{i},\mathbf{j}\right>} \left(\beta^{\dag}_\mathbf{i} \beta_\mathbf{j} + \beta^{\dag}_\mathbf{j}\beta_\mathbf{i}\right) + \sum_i \left(h-4J\right)\beta^{\dag}_\mathbf{i} \beta_\mathbf{i} + J\sum_{\left<\mathbf{i},\mathbf{j}\right>}\beta^{\dag}_\mathbf{i} \beta^{\dag}_\mathbf{j} \beta_\mathbf{i} \beta_\mathbf{j}.\label{eq:magnonBEC} \end{equation}
The $\beta_i$ and $\beta_j$ satisfy bosonic commutation relations for $\mathbf{i} \neq \mathbf{j}$. To avoid difficulty at $\mathbf{i}=\mathbf{j}$ we introduce an infinite on-site repulsion, or hard-core constraint, and can then regard $\beta^\dag_\mathbf{i}$ as the creation operator for a boson on site $\mathbf{i}$ \cite{SuppBatyev1984}. From the middle term in in Eq.~\ref{eq:magnonBEC}, we see the field $h$ is the chemical potential of the magnons (up to a constant offset). The last term describes nearest neighbor repulsion of magnons. The first term describes the bosons hopping. Transforming to momentum space, this term becomes $\sum_\mathbf{q}\epsilon(\mathbf{q})\beta^{\dag}_{\mathbf{q}}\beta_{\mathbf{q}}$ with $\epsilon(\mathbf{q}) = J \left( \cos(q_x a) + \cos(q_y a) \right)$ where $a$ is the lattice constant. In contrast to the typical case, the hopping term here is positive and the condensate forms at the band minimum, $\mathbf{q} = (\pi,\pi)$. Condensation is signalled by non-zero expectation value of off-diagonal density matrix elements. For system of volume $V$ and total particle number $N$ these elements are 
\begin{eqnarray*}
\left< \beta^{\dag}_\mathbf{i} \beta_{\mathbf{i}+\mathbf{d}} \right> &=& \frac{1}{V} \int d\,\mathbf{q} \exp\left(i \mathbf{q}\cdot \mathbf{d}/\hbar\right) \left<\beta^{\dag}_\mathbf{q} \beta_\mathbf{q}\right> \\
&=& \frac{N}{V}(-1)^{d_x+d_y}. 
\end{eqnarray*}
Using the identity $\beta^{\dag}_\mathbf{i}\beta_{\mathbf{i}+\mathbf{d}} + \beta^{\dag}_{\mathbf{i}+\mathbf{d}}\beta_\mathbf{i} = 2 \left(S^x_\mathbf{i}S^x_{\mathbf{i}+\mathbf{d}} + S^y_\mathbf{i}S^y_{\mathbf{i}+\mathbf{d}} \right)$, we see the BEC off-diagonal order is associated with spin correlations perpendicular to the magnetization, and the anti-ferromagnetic checkerboard is associated with the condensate quasimomentum, $(\pi,\pi)$. 

The spin correlations parallel to the magnetization are associated with the density correlations of the magnons, which can be written $\left<\beta^\dag_\mathbf{i} \beta_\mathbf{i} \beta^\dag_{\mathbf{i}+\mathbf{d}} \beta_{\mathbf{i}+\mathbf{d}} \right>_c = \left<S^z_\mathbf{i} S^z_{\mathbf{i}+\mathbf{d}} \right>_c \propto C^z(\mathbf{d})$. These are the density correlations of a liquid of bosons on the lattice, with hard core exclusion and nearest-neighbor repulsion.  These are expected to be typical liquid correlations, starting negative at short distances and oscillating and damping with distance, first becoming positive at a distance of order the inter-particle spacing.  As we go to high polarization and thus low density of these bosons, this inter-particle distance grows and moves beyond the diagonal neighbor distance, resulting in the sign change in $C^z(1,1)$.


\end{document}